\documentclass[aps,11pt,indentfirst,bm,preprint]{revtex4}
\usepackage{epsfig}
\usepackage{epsf}
\usepackage{amsmath}
\usepackage{amssymb}
\usepackage{slashed}
\newcommand{\PLB}[3]{Phys. Lett. B {\bf #1}, {#2} (#3)}
\newcommand{\PRL}[3]{Phys. Rev. Lett. {\bf #1}, {#2} (#3)}

\newcommand{\PRD}[3]{Phys. Rev. D {\bf #1}, {#2} (#3)}
\newcommand{\NPA}[3]{Nucl. Phys. {\bf A#1}, {#2} (#3)}
\newcommand{\NPB}[3]{Nucl. Phys. {\bf B#1}, {#2} (#3)}

\newcommand{\EPJC}[3]{Eur. Phys. J. C {\bf #1}, {#2} (#3)}
\newcommand{\ZPC}[3]{Z. Phys. C {\bf #1}, {#2} (#3)}
\newcommand{\JPG}[3]{J. Phys. G {\bf #1}, {#2} (#3)}
\newcommand{\PRt}[3]{Phys. Rept. {\bf #1}, {#2} (#3)}
\makeatletter
\def\underbracket{%
    \@ifnextchar[{\@underbracket}{\@underbracket [\@bracketheight]}%
}
\def\@underbracket[#1]{%
    \@ifnextchar[{\@under@bracket[#1]}{\@under@bracket[#1][0.4em]}%
}
\def\@under@bracket[#1][#2]#3{
    \mathop{\vtop{\m@th \ialign {##\crcr $\hfil \displaystyle {#3}\hfil $
    \crcr \noalign {\kern 3\p@ \nointerlineskip }\upbracketfill {#1}{#2}
     \crcr \noalign {\kern 3\p@ }}}}\limits}
\def\upbracketfill#1#2{$\m@th \setbox \z@ \hbox {$\braceld$}
                   \edef \@bracketheight{\the \ht \z@}\bracketend{#1}{#2}
                   \leaders \vrule \@height #1 \@depth \z@ \hfill
                   \leaders \vrule \@height #1 \@depth \z@ \hfill
                   \bracketend{#1}{#2}$}
\def\bracketend#1#2{\vrule height #2 width #1\relax}
\makeatother

\begin{document}


\title{Light vector hybrid states via QCD sum rules}

\author{Feng-Kun Guo$^{1,7}$}
\email{guofk@mail.ihep.ac.cn}
\author{Peng-Nian Shen$^{1,4,5}$}
\author{Zhi-Gang Wang$^2$}
\author{Wei-Hong Liang$^3$}
\author{L.S. Kisslinger$^6$} \affiliation{\small $^1$Institute of High Energy Physics,
Chinese
Academy of Sciences, P.O.Box 918(4), Beijing 100049, China\footnote{Corresponding address.}\\
$^2$Department of Physics, North China Electric Power University,
Baoding 071003, China\\
$^3$Department of Physics, Guangxi Normal University, Guilin 541004,
China\\
$^4$Institute of Theoretical Physics, Chinese Academy of Sciences, Beijing 100080, China\\
$^5$Center of Theoretical Nuclear Physics, National Laboratory of
Heavy Ion Accelerator, Lanzhou 730000, China\\
$^6$Department of Physics, Carnegie Mellon University, Pittsburgh, Pennsylvania 15213, USA\\
$^7$Graduate University of Chinese Academy of Sciences, Beijing
100049, China}

\date{\today}

\begin{abstract}
Vector hybrid states with light quarks $u,d,s$ are investigated via
QCD sum rules. The results show that the masses of the $q{\bar q}g$
$(q=u,d)$, $q{\bar s}g$, and $s{\bar s}g$ states with
$J^{PC}=1^{--}$ are about 2.3-2.4, 2.3-2.5, and 2.5-2.6 GeV,
respectively. It suggests that the recently discovered $Y(2175)$
could not be a pure $s{\bar s}g$ vector hybrid state.
\end{abstract}


\maketitle

It has been long expected that in QCD, besides the conventional
$q{\bar q}$ mesons and $qqq$ baryons, exotic states such as the
multiquark states and hybrid states, should exist as a consequence
of the non-perturbative aspect of QCD \cite{ja76,amsl}. A multiquark
state is composed of more than three quarks and anti-quarks; a
hybrid state contains valence gluon(s), besides valence quarks.
There has been lots of work on the hybrid states with exotic quantum
numbers, such as $J^{PC}=0^{--},~0^{+-},~1^{-+},~2^{+-}$ etc., which
cannot be obtained by using only a quark and an anti-quark
\cite{amsl}. In this paper, we will investigate the light flavor
hybrid states with quantum numbers $J^{PC}=1^{--}$.

In the present experimental spectrum of vector mesons, some states
are argued to be vector hybrid candidates or contain hybrid
components. For instance, the $\rho(1450)$ and $\omega(1420)$ were
proposed to have significant hybrid components because their decay
patterns are different from those expected for radially excited
$q{\bar q}$ states \cite{dk93,cd94,cp95,cp97}. The heavier state
$\omega(1600)$ was suggested to be $2S$-hybrid mixtures \cite{cp97}.
For the strange counterpart, $K^*(1410)$, both of its mass and decay
pattern have not been understood yet (\cite{bg97,bbp}). The
$K^*(1410)$ is too light to be a $2^3S_1$ state and its decay
pattern is also against the  $2^3S_1$ assignment \cite{bbp}. Its
largest decay channel is $K^*\pi$, with a branching fraction larger
than $40\%$; the branching fraction of the $K\pi$ channel is
$6.6\pm1.3\%$, and the one of the $K\rho$ channel is less than $7\%$
\cite{pdg06}. In Ref. \cite{bbp}, it was suggested that the low mass
of the $K^*(1410)$ state might be due to the presence of additional
hybrid mixing states.

The recent discovered charmonium-like vector state $Y(4260)$
\cite{4260} was suspected to be a $c{\bar c}g$ hybrid state
\cite{zh05}. Recently, a structure at about 2175 MeV was observed by
the BABAR Collaboration, and it is consistent with a resonance with
a mass of $M_X=2175\pm10\pm15$ MeV and width of
$\Gamma_X=58\pm16\pm20$ MeV \cite{2175}. It was observed in
$e^+e^-\to \phi f_0(980)$ via initial state radiation, hence has
quantum numbers of $J^{PC}=1^{--}$. Immediately, a suggestion of an
$ss{\bar s}{\bar s}$ tetraquark state appeared \cite{wang}. Such an
interpretation was criticized for its possible large width, and then
an $s{\bar s}g$ hybrid interpretation was proposed \cite{dy06}. It
was pointed out in Ref. \cite{ro06} that the hybrid suggestion would
make sense if the $Y(4260)$ is a $c{\bar c}$ hybrid and
$m_c-m_s\simeq (M_Y-M_X)/2=1.04$ GeV.

To understand all the states mentioned above, an important step is
to calculate the masses of vector hybrid states. Isgur and Paton
estimated the masses of the $q{\bar q}g$ ($q=u,d$), $n{\bar s}g$ and
$s{\bar s}g$ states to be about 1.9, 2.0 and 2.1 GeV, respectively,
by using the flux-tube model \cite{ip85}. However, they stated that
the reliability was no better than $\pm100$ MeV, and the
spin-dependent perturbations were not considered. In this paper, we
study the light flavor hybrid state with $J^{PC}=1^{--}$ by using
the method of QCD sum rules (QCDSR) which has been proved to be very
successful in many hadronic problems \cite{svz,rev}. The central
idea of QCDSR is to calculate the correlation function of an
interpolating current with definite $J^{PC}$ quantum numbers from
both the phenomenological side ($q^2>0$) and the QCD side $q^2<<0$
through operator product expansion (OPE). The expression on the
phenomenological side can be related to that on the QCD side via
dispersion relation to get sum rules, and then hadronic parameters
can be determined.

We start from the two-point correlation function of the
interpolating vector current $J_{\mu}(x)$
\begin{eqnarray}
\Pi_{\mu\nu}(q^2) &\!=&\! i\int d^4x e^{iq\cdot
x}\langle0|T\{J_{\mu}(x)J_{\nu}^{\dag}(0)\} |0\rangle \nonumber\\
&\!=&\! (-g_{\mu\nu}+\frac{q_{\mu}q_{\nu}}{q^2})\Pi_1(q^2) +
\frac{q_{\mu}q_{\nu}}{q^2}\Pi_0(q^2).
\end{eqnarray}
The interpolating current for a vector hybrid state with quantum
numbers $J^{PC}=1^{--}$ is taken as
\begin{equation}
J_{\mu}(x) = g_s{\bar \psi}_A^a(x) \gamma^{\nu} \gamma_5
\frac{\lambda^n_{ab}}{2} \tilde{G}^n_{\mu\nu}(x) \psi_B^b(x),
\end{equation}
where $g_s$ is the strong coupling constant, $A,B=u,d,s$ and
$a,b=1,2,...,8$ are flavor and color indices, respectively.
$\tilde{G}^n_{\mu\nu}(x)=\varepsilon_{\mu\nu\alpha\beta}G^{n,\alpha\beta}(x)/2$
is the dual field strength of $G^n_{\mu\nu}(x)$. The overlapping
amplitude of this current with the vector hybrid state $X$ is
defined as
\begin{equation}
\langle 0|J_{\mu}(0)|X\rangle = f_X m_X^3 \epsilon_{\mu},
\end{equation}
where $f_X$, $m_X$ and $\epsilon_{\mu}$ are the decay constant, mass
and polarization vector of the hybrid state, respectively.

For the phenomenological side, we can obtain the imaginary part of
the correlation function as
\begin{equation}
\frac{\textrm{Im}\Pi(q^2)}{\pi}=f_X^2m_X^6\delta(q^2-m_X^2)+\rho^h(q^2)\theta(q^2-s_0^h)
\end{equation}

From the dispersion relation without subtractions (because
subtractions will be removed by the Borel transform, we can neglect
them here), we have
\begin{equation}
\Pi(q^2) = \frac{f_X^2m_X^6}{m_X^2-q^2-i\epsilon} +
\int_{s_0}^{\infty}\frac{\rho^h(s)}{s-q^2-i\epsilon}ds,
\end{equation}
where $\rho^h(s)$ represents the spectral function of the continuum
states, and $s_0$ is the threshold parameter.

\begin{figure}[hbt]
\begin{center}\vspace*{0.5cm}
{\epsfysize=8cm \epsffile{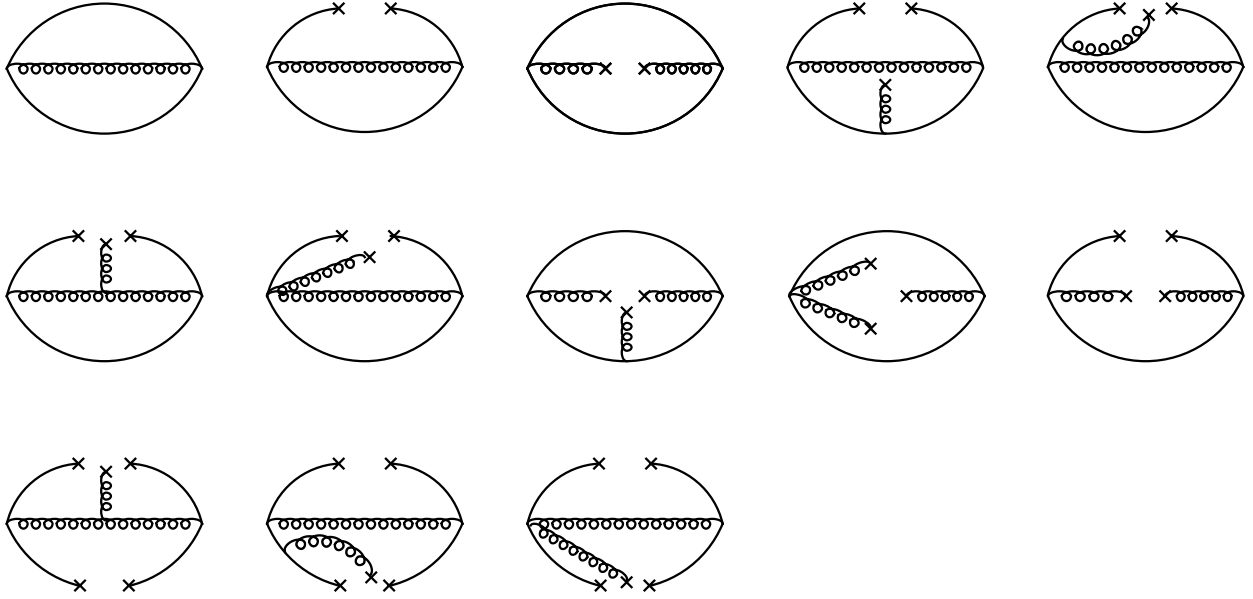}}%
\vglue -0.cm\caption{\label{fig:feyn} The relevant Feynman diagrams
for calculating the sum rules for light vector hybrid states.
Permutation diagrams are implied and not shown here.}
\end{center}
\end{figure}

The correlation function can be treated in the framework of the
operator product expansion (OPE), then the short and long distance
interactions are separated. The short distance interactions are
encoded in the Wilson coefficients which can be calculated by using
perturbative QCD at large $Q^2=-q^2$, and the long distance effects
are parameterized in a sets of universal quark and gluon condensates
\cite{svz}. When performing operator product expansion (OPE), we
consider operators up to eight dimension. The relevant Feynman
diagrams for deriving the QCD sum rules for light vector hybrid
states are shown in Fig. \ref{fig:feyn}.

After tedious calculations, the correlation function of the $s{\bar
s}g$ vector hybrid is obtained as
\begin{eqnarray}
\Pi_{s{\bar s}g}(p^2) &\!\!=&\!\! \left(
-\frac{\alpha_s}{240\pi^3}p^6 + \frac{5\alpha_s}{48\pi^3}m_s^2p^4
-\frac{4\alpha_s}{9\pi}m_s\langle {\bar s}s\rangle p^2 +
\frac{1}{144\pi^2}\langle g_s^2 G^2\rangle p^2 -
\frac{1}{16\pi^2}m_s^2\langle g_s^2 G^2\rangle \right. \nonumber\\
&\!\!&\!\! \left.  - \frac{49\alpha_s}{144\pi}m_s\langle g_s{\bar
s}Gs\rangle - \frac{1}{64\pi^2}\langle g_s^3G^3\rangle \right)
\ln{(-p^2)} \nonumber\\
&\!\!&\!\! + \frac{m_s^2}{32\pi^2p^2}\langle g_s^3G^3\rangle +
\frac{m_s}{12p^2} \langle {\bar s}s\rangle \langle g_s^2G^2\rangle -
\frac{\pi\alpha_s}{36p^2} \langle {\bar s}s\rangle \langle g_s{\bar
s}Gs\rangle,
\end{eqnarray}
where $\langle g_s{\bar s}Gs\rangle= \langle g_s{\bar
s}\sigma^{\mu\nu}t^nG^n_{\mu\nu}s\rangle$ with $t^n=\lambda^n/2$
being SU(3) generators, $\langle g_s^3G^3\rangle=\langle
g_s^3f^{lmn} G_{\gamma\delta}^l G_{\delta\epsilon}^m
G_{\epsilon\gamma}^n\rangle$. Similarly, the correlation functions
of the $q{\bar s}g$, and $q{\bar q}g$ ($q=u,d$) vector hybrid states
are
\begin{eqnarray}
\Pi_{q{\bar s}g}(p^2) &\!\!=&\!\! \left( -
\frac{\alpha_s}{240\pi^3}p^6  + \frac{3\alpha_s}{64\pi^3}m_s^2p^4 -
\frac{\alpha_s}{18\pi}m_s\langle {\bar q}q\rangle p^2
-\frac{\alpha_s}{6\pi}m_s\langle {\bar s}s\rangle p^2 +
\frac{1}{144\pi^2}\langle g_s^2 G^2\rangle p^2 \right.\nonumber\\
&\!\!&\!\! \left. - \frac{1}{64\pi^2}m_s^2\langle g_s^2 G^2\rangle +
\frac{3\alpha_s}{64\pi}m_s\langle g_s{\bar q}Gq\rangle +
\frac{\alpha_s}{192\pi}m_s\langle g_s{\bar s}Gs\rangle
- \frac{1}{64\pi^2}\langle g_s^3G^3\rangle \right) \ln{(-p^2)} \nonumber\\
&\!\!&\!\!  + \frac{5m_s^2}{384\pi^2p^2}\langle g_s^3G^3\rangle +
\frac{m_s}{24p^2} \langle {\bar q}q\rangle\langle g_s^2G^2\rangle -
\frac{\pi\alpha_s}{72p^2} \langle {\bar q}q\rangle \langle g_s{\bar
s}Gs\rangle - \frac{\pi\alpha_s}{72p^2} \langle {\bar
s}s\rangle \langle g_s{\bar sq}Gq\rangle,\\
\Pi_{q{\bar q}g}(p^2) &\!\!=&\!\! \left(
-\frac{\alpha_s}{240\pi^3}p^6 + \frac{1}{144\pi^2}\langle g_s^2
G^2\rangle p^2 - \frac{1}{64\pi^2}\langle g_s^3G^3\rangle \right)
\ln{(-p^2)} - \frac{\pi\alpha_s}{36p^2} \langle {\bar q}q\rangle
\langle g_s{\bar q}Gq\rangle,
\end{eqnarray}
respectively.

The Borel transform are defined as
\begin{equation}
\hat{{\cal B}}_{M_B^2}f(p^2) =
\lim_{\substack{-p^2,n\to\infty\\-p^2/n=M_B^2}}
\frac{(-p^2)^{n+1}}{n!} \left( \frac{d}{dp^2} \right)^n f(p^2).
\end{equation}
Performing the Borel transform to both the phenomenological side and
the QCD side, and using the quark-hadron duality to approximate the
continuum contribution, we obtain the following sum rules
\begin{eqnarray}
f_{s{\bar s}g}^2m_{s{\bar s}g}^6 e^{-m_{s{\bar s}g}^2/M_B^2}
&\!\!=&\!\! \int_0^{s_0}dp^2 \left( \frac{\alpha_s}{240\pi^3}p^6 -
\frac{5\alpha_s}{48\pi^3}m_s^2p^4 +\frac{4\alpha_s}{9\pi}m_s\langle
{\bar s}s\rangle p^2 - \frac{1}{144\pi^2}\langle g_s^2 G^2\rangle
p^2 \right. \nonumber\\
&\!\!&\!\! \left.  + \frac{1}{16\pi^2}m_s^2\langle g_s^2 G^2\rangle
+ \frac{49\alpha_s}{144\pi}m_s\langle g_s{\bar s}Gs\rangle +
\frac{1}{64\pi^2}\langle g_s^3G^3\rangle \right)
e^{-p^2/M_B^2} \nonumber\\
&\!\!&\!\! - \frac{m_s^2}{32\pi^2}\langle g_s^3G^3\rangle -
\frac{m_s}{12} \langle {\bar s}s\rangle \langle g_s^2G^2\rangle +
\frac{\pi\alpha_s}{36} \langle {\bar s}s\rangle \langle g_s{\bar
s}Gs\rangle,\\
f_{q{\bar s}g}^2m_{q{\bar s}g}^6 e^{-m_{q{\bar s}g}^2/M_B^2}
&\!\!=&\!\! \int_0^{s_0}dp^2 \left( \frac{\alpha_s}{240\pi^3}p^6 -
\frac{3\alpha_s}{64\pi^3}m_s^2p^4 + \frac{\alpha_s}{18\pi}m_s\langle
{\bar q}q\rangle p^2 +\frac{\alpha_s}{6\pi}m_s\langle {\bar
s}s\rangle p^2 -
\frac{1}{144\pi^2}\langle g_s^2 G^2\rangle p^2 \right.\nonumber\\
&\!\!&\!\! \left. + \frac{1}{64\pi^2}m_s^2\langle g_s^2 G^2\rangle -
\frac{3\alpha_s}{64\pi}m_s\langle g_s{\bar q}Gq\rangle -
\frac{\alpha_s}{192\pi}m_s\langle g_s{\bar s}Gs\rangle
+ \frac{1}{64\pi^2}\langle g_s^3G^3\rangle \right) e^{-p^2/M_B^2} \nonumber\\
&\!\!&\!\!  - \frac{5m_s^2}{384\pi^2}\langle g_s^3G^3\rangle -
\frac{m_s}{24} \langle {\bar q}q\rangle\langle g_s^2G^2\rangle +
\frac{\pi\alpha_s}{72} \langle {\bar q}q\rangle \langle g_s{\bar
s}Gs\rangle + \frac{\pi\alpha_s}{72} \langle {\bar
s}s\rangle \langle g_s{\bar sq}Gq\rangle,\\
f_{q{\bar q}g}^2m_{q{\bar q}g}^6 e^{-m_{q{\bar q}g}^2/M_B^2}
&\!\!=&\!\! \int_0^{s_0}dp^2 \left( \frac{\alpha_s}{240\pi^3}p^6 -
\frac{1}{144\pi^2}\langle g_s^2 G^2\rangle p^2 +
\frac{1}{64\pi^2}\langle g_s^3G^3\rangle \right) e^{-p^2/M_B^2}
\nonumber\\&\!\!&\!\! + \frac{\pi\alpha_s}{36} \langle {\bar
q}q\rangle \langle g_s{\bar q}Gq\rangle,
\end{eqnarray}

The running coupling constant can be taken as $\alpha_s =
4\pi/\left[(11-2/3n_f)\ln{(M_B^2/\Lambda^2_{QCD})}\right]$ with
three active flavors. For numerical analysis, we use the following
inputs: \cite{svz,jo06,yh93,gl05}
\begin{eqnarray}
\Lambda_{QCD} &\!\!=&\! 220 ~\textrm{MeV},\nonumber\\
m_s(2\textrm{GeV}) &\!\!=&\! 94 ~\textrm{MeV},\nonumber\\
\langle {\bar q}q \rangle &\!\!=&\! -(0.024~\textrm{GeV})^3, \nonumber\\
\langle {\bar s}s \rangle &\!\!=&\! 0.8\times\langle {\bar q}q \rangle, \\
\langle g_s^2G^2\rangle &\!\!=&\! 0.48 ~\textrm{GeV}^4, \nonumber\\
\langle g_s{\bar q}Gq\rangle &\!\!=&\! m_0^2\langle {\bar q}q
\rangle ~\textrm{with}~m_0^2=0.8~\textrm{GeV}^2, \nonumber\\
\langle g_s^3G^3\rangle &\!\!=&\! 0.045 ~\textrm{GeV}^6. \nonumber
\end{eqnarray}

According to Ref. \cite{gv84}, the physical information of $m_X$ and
$f_X$ can be extracted by fitting the right hand side and the left
hand side with $m_X^2$ and $f_X^2$ as free parameters. The $\chi^2$
fit is done in a reasonable interval of values of $M_B^2$ to
guarantee that the contributions of the operators from dimension 8
are less than 10\%, and those of the continuum are less than 50\%.
The results are given in Table \ref{tab:results}.
\begin{table}[htb]
\caption{\label{tab:results} The values of $f_X$, $m_X$, and
$\sqrt{s_0}$ from fitting the left hand side to the right hand
side.}
\begin{ruledtabular}
\begin{tabular}{lccc}
 & $m_X$ (GeV) & $f_X$ (MeV) & $\sqrt{s_0}$ (GeV) \\ \hline%
$q{\bar q}g$ & 2.33 & 10.8 & 2.6 \\
             & 2.34 & 12.3 & 2.8 \\%
             & 2.43 & 13.9 & 3.0 \\
$q{\bar s}g$ & 2.34 & 11.0 & 2.7 \\%
             & 2.41 & 12.6 & 2.9 \\
             & 2.50 & 14.2 & 3.1 \\
$s{\bar s}g$ & 2.54 & 11.3 & 2.8 \\%
             & 2.54 & 12.8 & 3.0 \\
             & 2.62 & 14.3 & 3.2 \\
\end{tabular}
\end{ruledtabular}
\end{table}
The results show that the masses of the $1^{--}$ hybrid states are
well above 2 GeV. Compared with the results from the flux tube model
\cite{ip85}, the values obtained here are 400-500 MeV higher.

In this paper, we calculate the masses and decay constants of the
light flavor $1^{--}$ hybrid states. The masses of all the $q{\bar
q}g$, $q{\bar s}g$ and $s{\bar s}g$ vector states are above 2 GeV,
being about 2.3-2.4, 2.3-2.5, and 2.5-2.6 GeV, respectively. From
these results, the mesons $\rho(1450)$, $\omega(1420)$,
$\omega(1600)$, $K^*(1410)$, and etc. would not be pure hybrid
states, but they could contain hybrid mixtures. compared with these
states, the recently discovered vector state $Y(2175)$ is stated
closer to the position of the predicted hybrid state, $s{\bar s}g$.
However, its mass is still lower than the one obtained here, which
imply that the $Y(2175)$ state could not be a pure hybrid state.
This state could be an excited $s{\bar s}$ state \cite{bugg} or its
mixing with a hybrid state. Both effort from the experimental side
and the theoretical side are necessary to identify the nature of the
$Y(2175)$ state. Experimentally, other decay channels of the state
should be hunted out; theoretically, decay properties of the vector
$s{\bar s}g$ hybrid state and the $2^3D_1$ $s{\bar s}$ state with
mass of 2175 MeV have been investigated by using the flux tube model
and the $^3P_0$ model \cite{dy06,dy07}. By using the information on
the mass of vector hybrid states, possible mixing between the
$s{\bar s}g$ and the $2^3D_1$ $s{\bar s}$ could be studied.

\begin{acknowledgments}
We sincerely acknowledge D.V. Bugg for useful discussion on the
$Y(2175)$ state. This work is partially supported by the NSFC grant
Nos. 90103020, 10475089, 10435080, 10447130, CAS Knowledge
Innovation Key-Project grant No. KJCX2SWN02 and Key Knowledge
Innovation Project of IHEP, CAS (U529). One of the authors (WZG)
would like to thanks National Natural Science Foundation,Grant
Number 10405009, for financial support.
\end{acknowledgments}

\end{document}